\documentclass{elsart}
\usepackage{natbib}
\usepackage{graphicx}
\usepackage{amssymb}

\begin{document}

\begin{frontmatter}



\title{The value of the equation of state of dark energy}


\author{Raul Jimenez} 
\ead{raulj@physics.upenn.edu}
\address{Department of Physics and Astronomy, University of
Pennsylvania, Philadelphia, PA 19104, USA}

\begin{abstract}
From recent CMB and Large Scale Structure observations the value of
the equation of state of dark energy, assuming it to be constant in
time, is constrained to be $-1.3<w<-0.7$ at the $95\%$ confidence
level: consistent with dark energy being a classical vacuum term.
Here we describe two novel and independent methods, sensitive to
different systematics, that give the same value for $w$ and similar
confidence regions. This suggests that systematics are not an issue in
current determinations of $w$. The first method yields a measurement
of $w$ that relies on the minimum number of model-dependent
parameters; the second method is a non-parametric measurement of the
time dependence of $w(z)$. We also present a method to statistically
determine the edge of a distribution.
\end{abstract}

\begin{keyword}


\end{keyword}

\end{frontmatter}


\section{Introduction}

It is now established that the universe is accelerating (e.g.,
\citet{D+96,Spinrad+97,Riess+98,Perlmutter+99}): the next logical step
is to unveil the nature of the accelerating force, for example, by
determining its equation of state ($p=w(z) \rho$), where $w(z)$ need
not to be constant and can vary with redshift.

During the past year we have seen a significant improvement in the
accuracy of the equation of state, $w$ measurements (assuming $w$ to
be constant in time, e.g. \citet{SpergelWMAP03,CKW03,JVTS03,Tonry+03}
and also Verde and Melchiorri articles in these proceedings). All
these methods are sensitive to different systematics, yet the results
are remarkably consistent: $ -1.3 < w < -0.7$ at 95\% confidence
\citep{SpergelWMAP03}. While this measurement will soon become even
tighter with 2yr WMAP data and SDSS galaxy and Ly$-\alpha$ forest
power spectrum release, it still assumes $w$ to be constant in time. A
significant challenge will be to accurately measure $w(z)$. There has
been a number of papers discussing different methods to measure $w$
(e.g. \citet{AlcockPaczynski79,TurnerWhite97,CDS98,Garnavich+98,Birkinshaw99,Efstathiou99,Hui99,NewmanDavis00,HaimanMohrHolder01,HutererTurner01,WangGarnavich01,WellerAlbrecht01,Baccigalupi+02,Huterer02,Hu02,KLSW02,Hu03,LimaAlcaniz02})
while less attention has been devoted to the more challenging task of
measuring $w(z)$ without parameterizations.

Below, we describe two novel methods: a) a measurement of $w$ that
relies on the minimum number of model dependent parameters. This
method uses the determination of the position of the first acoustic
peak in the CMB angular power spectrum ($\ell_1$) and the absolute ages
of galactic globular clusters (GCs). b) a non-parametric measurement
of the time dependence of $w(z)$. This method is based on the {\em
relative} ages of stellar populations. These techniques have been
described in \citet{JL02, JVTS03}

\section{$w$ from the $\ell_1-$GCs ages method}

If the universe is assumed to be flat, the position of the first
acoustic peak ($\ell_1$ in the standard spherical harmonics notation)
depends primarily on the age of the universe and on the effective
value\footnote{I.e., the average value over redshift. We use
$w(z)$ to indicate when $w$ is allowed to vary over time.} of $w$
\citep{CDS98,HFZT01,KCS01,CKW03}.  As noted by these authors, for a fixed $w$
value, a change in the physical density parameter $\omega_m=\Omega_mh^2$
that keeps the characteristic angular scale of the first acoustic peak
fixed will also leave the age approximately unchanged. Thus an independent
estimate of the absolute age of the universe at $z=0$ combined with a
measurement of $\ell_1$ yields an estimate of $w$ largely independent
of other cosmological parameters (e.g., $\Omega_m$ and $h$).

This can be better understood by considering that for a constant $w(z)$,
the age of a flat universe is given by
\begin{equation}
t_0= H_0^{-1} \int_{0}^{\infty} (1+z)^{-1}\left[\Omega_{\rm
rad}(1+z)^4+\Omega_m(1+z)^3+\Omega_{\Lambda} (1+z)^{3(1+w)}\right]^{-1/2}
dz\;.
\label{eq:ageflatwconst}
\end{equation}
The position of the first acoustic peak is fixed by the quantity
\begin{equation}
\theta_A=r_s(a_{dec})/d_A(a_{dec})\;,
\label{eq:thetaa}
\end{equation}
where $a_{dec}$ is the scale factor at decoupling, $r_s(a_{dec})$
is the sound horizon at decoupling, and $d_A(a_{dec})$ is the angular
diameter distance at decoupling.  For a flat universe,
\begin{equation}
r_s(a_{dec})=\frac{c}{H_0\sqrt{3}}\int_0^{a_{dec}}\left[\left(1 +
\frac{3\Omega_b} {4\Omega_{\gamma}}\right)(\Omega_{\Lambda}x^{1-3w} +
\Omega_mx + \Omega_{\rm rad})\right]^{-1/2}dx
\label{eq:rs}
\end{equation}
and
\begin{equation}
d_A(a_{dec})=\frac{c}{H_0}\int_{a_{dec}}^1\left[\Omega_{\Lambda}x^{1-3w}
+ \Omega_mx + \Omega_{\rm rad}\right]^{-1/2}dx\;.
\label{eq:dA}
\end{equation}
See \citet{VerdeWMAP03} for more details. We use equations
\ref{eq:ageflatwconst} to \ref{eq:dA} and fix $\theta_A$ to be
consistent with the {\it WMAP} determination \citep[$\ell_1\simeq
220$;][]{PageWMAP03}. Figure~1 shows the allowed region in the
age-$w$ plane obtained for $\Omega_b h^2=0.02$, $0.05<\Omega_m<0.4$,
and $0.5<h<0.9$. It is clear from the plot that an independent and
accurate age determination of the universe will provide a measurement
of $w$. Thus by using only the {\it WMAP} observation of $\ell_1$ and
an independent estimate of the age of the universe, one can place
constraints on $w$, largely independent of other cosmological
parameters. Note that this method is sensitive to a different redshift
weighting than supernova and CMB measurements.

\begin{figure}
\begin{center}
\includegraphics*[width=10cm]{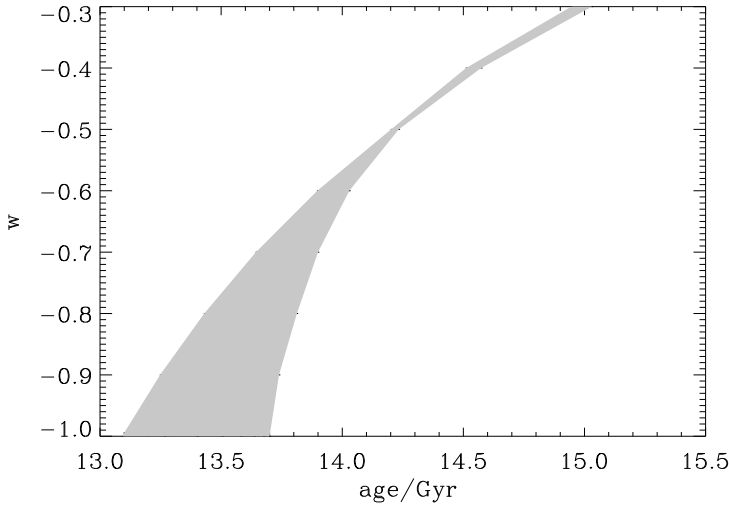}
\end{center}
\caption{The region of parameter space allowed if we fix $\theta_A$
    (which in turns fixes the position of the first CMB acoustic peak)
    and $\omega_b \equiv \Omega_b h^2 = 0.02$, weakly constraining
    $0.05<\Omega_m<0.45$ and $0.5<h<0.9$.}
\label{fig:cmb-w}
\end{figure}

The ages of the oldest globular clusters (GCs) provide a lower limit
to the total age of the universe. Since numerous star-forming galaxies
have now been observed up to redshift $z=6.6$
\citep[e.g.,][]{Kodaira+03}and the oldest GCs contain the oldest
stellar populations in galaxies, it is reasonable to assume that GCs
too have formed at redshift $z>6$.  \citet{KraussChaboyer03} perform
the most careful analysis to date of the effects of systematics in GC
age determinations.  They estimate the age of the oldest Galactic GCs
using the main-sequence turn-off luminosity and evaluate the errors
with Monte-Carlo techniques, paying careful attention to uncertainties
in the distance, to systematics, and to model uncertainties.  They
find a best-fit age for the oldest Galactic GCs of
$12.5^{+3.4}_{-2.2}$~Gyr (95\% confidence limits). We find that their
probability distribution for the oldest GC age, can be accurately
described by
\begin{equation}
P(t)=\frac{A}{\sigma (t-T)} \exp \left[- \frac{\ln \left[ (t-T)/m \right
]^2}{2 \sigma^2} \right ]
\label{eq:kraussagecorr}
\end{equation}
where $t$ denotes the age of the oldest GCs, $A=1.466$, $\sigma=0.25$,
$T=6.5$ and $m=5.9$. Since the oldest GCs likely formed at $z>5$, for
all reasonable cosmologies we only need add about $0.3$ Gyr to the GC
ages to obtain an estimate of the age of the universe. Only the oldest
(low metallicity) GCs should be used in the age estimate, for which
there is no age spread (see e.g. \citet{Rosenberg+99}, which finds no
age spread for GCs with metallicities $[Fe/H] < -1.2$).

In order to constrain $w$, we can now assume a flat universe and Monte
Carlo simulate the distribution of $\ell_1$ subject to only weak
priors on the other cosmological parameters.  For different values of
$w$, \citet{JVTS03} generated $10^5$ models, randomly sampling the
cosmological parameters $\Omega_m$, $\Omega_bh^2$ and $h$ with uniform
priors, $0.5< h < 0.9$, $0.05 < \Omega_m < 0.45$, and $0.01 <
\Omega_bh^2 < 0.03$. The next step is to impose an age of the universe
constraint by randomly sampling these models so that the age of the
universe has a probability distribution whose shape is given by
equation \ref{eq:kraussagecorr}, but offset by $0.3$
Gyr. \citet{JVTS03} used the publicly available code {\sf CMBFAST}
\citep{SeljakZaldarriaga96} to compute the resulting distribution of
$\ell_1$. This is shown in Figure~2 where the two solid lines are the
68\% and 90\% {\it joint} confidence levels.

As expected, the age alone does not constrain $w$; there is a degeneracy
between $w$ and $\ell_1$.  If we now use the {\em observed} position
of the first acoustic peak as recently measured, in a model independent
way, from {\it WMAP} \citep{PageWMAP03}, we can constrain $w$ with high
accuracy. The filled contours in Figure~2 are marginalized values for $w$
at the 68\% and 90\% confidence levels. Thus we find $w<-0.8$ ($w < -0.67$)
at the 68\% (90\%) confidence level.  

This determination depends solely on the GC determination of the age
of the universe and on the {\em observed} position of the first
acoustic peak in the CMB power spectrum. This constraint is slightly
less stringent than that obtained by \citet{SpergelWMAP03} from a
joint likelihood analysis of {\it WMAP} with six external data sets
({\it WMAP} + CBI + ACBAR + 2dFGRS + Lyman$\alpha$ forest power
spectrum + Type~Ia supernovae + $H_0$ constraint from the {\it HST}
key project), but is tighter than the CMB-only ({\it WMAP} + CBI +
ACBAR) determination and comparable to the {\it WMAP} + ACBAR + CBI +
{\it HST} constraint.

\begin{figure}
\begin{center}
\includegraphics*[width=10cm]{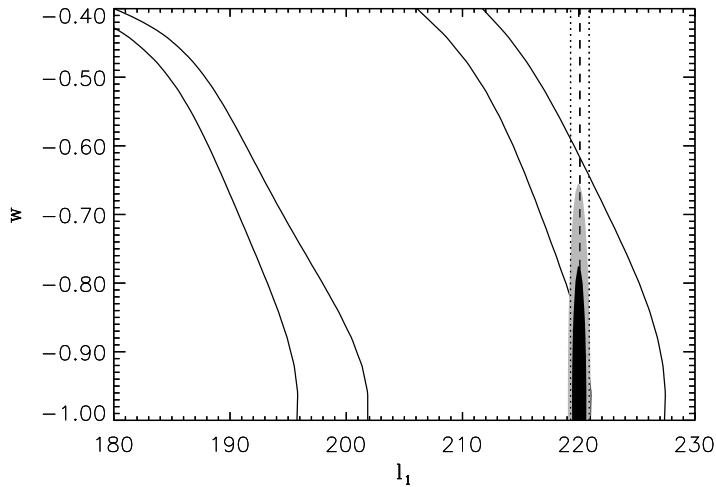}
\end{center} 
\caption{Confidence regions (68 and 90\%) for the value of $w$ are
    shown by the filled grey areas (marginalized over the position of
    the first acoustic peak in the CMB, $\ell_1$).  The solid lines
    denote contours which show the constraints in $w$ using the
    derived age of the oldest Galactic globular clusters only. The
    dash and dotted lines are the the position of the first acoustic
    peak and confidence region as measured by {\it WMAP}. Note that
    the position of the first peak greatly constrains the value of
    $w$.}
\label{lwplane}
\end{figure}

\section{Measuring a variable equation of state for the dark energy}

The popular approach for measuring $w(z)$ uses its effect on the
luminosity distance of sources.  In particular, the proposal for the
Supernova/Acceleration Probe (SNAP) mission\footnote{http://snap.lbl.gov/}
suggests to monitor $\sim 2000$ Type Ia supernovae across the sky per year
and determine their luminosity distances up to a redshift $z\sim 1.5$ with
high precision.  However, the sensitivity of the luminosity distance to the
redshift evolution of $w(z)$ is compromised by its integral nature (Maor et
al. 2001),
\begin{equation}
d_{\rm L}= (1+z)\int_z^0 (1+z^\prime){dt\over dz^\prime} dz^\prime ,
\label{eq:d_l}
\end{equation}
where $t(z)$ is the age of the Universe at a redshift $z$ which depends on
$w(z)$.

\citet{JL02} proposed an alternative method that offers a much better
sensitivity to $w(z)$ since it measures the integrand of
equation~(\ref{eq:d_l}) directly. Any such method must rely on a clock
that dates the variation in the age of the Universe with redshift.
The clock is provided by spectroscopic dating of galaxy ages. Based on
measurements of the age difference, $\Delta t$, between two
passively--evolving galaxies that formed at the same time but are
separated by a small redshift interval $\Delta z$, one can infer the
derivative, $(dz/dt)$, from the ratio $(\Delta z/\Delta t)$.  The
statistical significance of the measurement can be improved by
selecting fair samples of passively--evolving galaxies at the two
redshifts and by comparing the upper cut-off in their age
distributions. All selected galaxies need to have similar
metallicities and low star formation rates (i.e. a red color), so that
the average age of their stars would far exceed the age difference
between the two galaxy samples, $\Delta t$.

This {\it differential age method} is much more reliable than a method
based on an absolute age determination for galaxies (e.g., Dunlop et
al. 1996; Alcaniz \& Lima 2001; Stockton 2001). As demonstrated in the case
of globular clusters, absolute stellar ages are more vulnerable to
systematic uncertainties than relative ages (Stetson, Vandenberg \& Bolte
1996).  Moreover, absolute galaxy ages can only provide a lower limit to
the age of the Universe and only place weak constraints on the possible
histories of $w(z)$.

Consider a flat universe composed of matter and dark energy with an
equation of state $p_Q=w(z)\rho_Q$. The Hubble parameter is
$H^2=H_0^2[\rho_T(z)/\rho_T(0)]$.  Here, the subscripts $Q$, $m$, and
$T$ refer to the dark energy, the matter, or the total sum of the two,
respectively.  Assuming further that the matter is non-relativistic
(i.e. effectively pressureless), we get
\begin{equation}
H_0^{-1}{dz\over dt} = -(1+z)^{5/2} \left
[\Omega_m(0) +\Omega_Q(0) \exp \left \{3 \int_{0}^{z}
\frac{dz^\prime}{(1+z^\prime)} w(z) \right \} \right ]^{1/2} ,
\label{eq0}
\end{equation}
where we have used the energy conservation equation for the dark energy,
$\dot \rho_Q=-3H(1+w(z))\rho_Q$ (Maor et al. 2000).  Thus, $({dz}/{dt})$ is
related to the equation of state of the dark energy through one integration
only, while the luminosity distance in equation (\ref{eq:d_l}) is given by
an integral of the inverse of equation (\ref{eq0}), namely through two
integrations. By differentiating $(dz/dt)$ with respect to $t$ we find,
\begin{equation}
H_0^{-2}\frac{d^2z}{dt^2}=\frac{[H_0^{-1}({dz}/{dt})]^2}{(1+z)}\left
  [\frac{5}{2}+\frac{3}{2}w(z)
  \right]-\frac{3}{2}\Omega_m(0)(1+z)^4w(z) ,
\label{eq2}
\end{equation}
which depends {\it explicitly} on $w(z)$ without any integrations.  Thus,
the second derivative of redshift with respect to cosmic time measures
$w(z)$ directly. While it is possible to find significantly different
redshift histories of $w(z)$ for which the evolution of $d_{\rm L}$ is
similar, this cannot be done for $({d^2z}/{dt^2})$.

To obtain $dz/dt$, \citet{JL02} proposed to use the old envelope of
the age--redshift relation of E/S0 galaxies. They assume that E/S0 at
different redshifts are drawn from the same parent population with the
bulk of their stellar populations formed at relatively high redshift
\citep[e.g.,][]{Bower+92, Stanford+98}.  Then, at relatively low redshift,
they are evolving passively and may be used as ``cosmic chronometers''.

Recently, \citet{JVTS03} applied this method using high-quality
spectra of a sample of old stellar populations covering the largest
possible range in redshift. To this aim they combined the following
data sets: (i) the luminous red galaxy (LRG) sample from the Sloan
Digital Sky Survey (SDSS) early data release \citep{Eisenstein+01};
(ii) the sample of field early-type galaxies from \citet[][hereafter
the Treu et al.\ sample]{TSCMB99,TSMCB01,TSCMB02}; (iii) a sample of
red galaxies in the galaxy cluster MS1054$-$0321 at $z=0.83$; and (iv)
the two radio galaxies 53W091 and 53W069 \citep[][Dey et al., in
preparation]{D+96,Spinrad+97,NDJH03}. They obtained the age of the
dominant stellar population in the galaxies by fitting single stellar
population models \citep{JPMH98} to the observed spectrum.

\begin{figure}
\begin{center}
\includegraphics*[width=10cm]{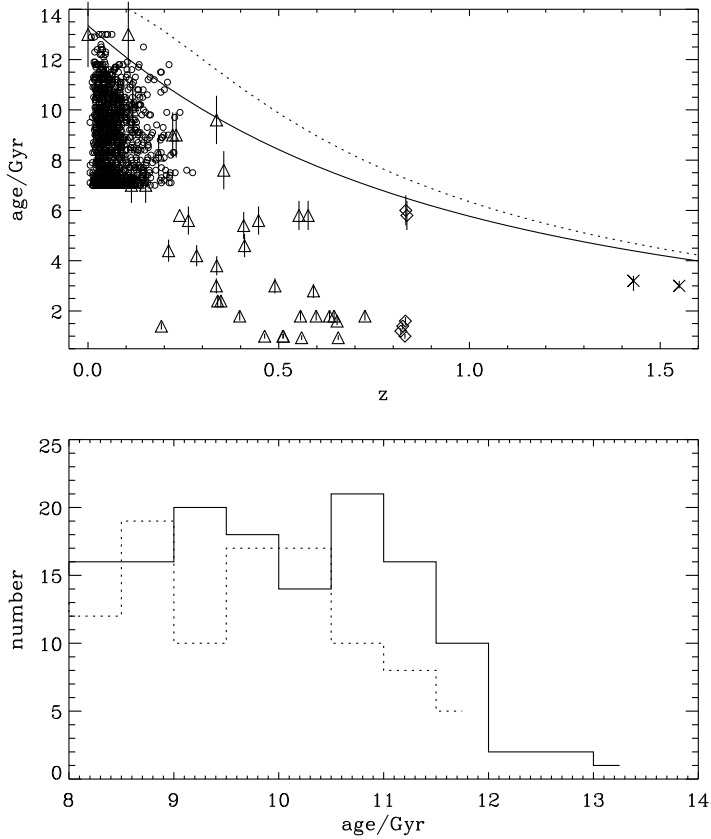}
\end{center} 
\caption{Top panel: Age--redshift envelope obtained for the galaxies
    studied in \citet{JVTS03}. Open circles correspond to galaxies from
    the SDSS LRG sample, triangles to the Treu et al.\ sample,
    diamonds to MS1054$-$0321, and crosses are 53W091 and 53W069.  A
    clear trend is present: galaxies age as the redshift decreases.
    The overall shape of this trend is in fair agreement with
    theoretical expectations for a $\Lambda$CDM (solid line) with
    $\Omega_m=0.27$ and $H_0=71$ km s$^{-1}$ Mpc$^{-1}$.  The dotted
    line represent an alternative model disfavored by the data:
    $w(z)=-2$ for $z > 1$ and then grows linearly from $w=-2$ at $z=1$
    to $w=0$ at $z=0$. Bottom panel: histogram along the age axis of
    top panel for the redshift range $0 < z < 0.04$ and $0.08 < z <
    0.12$. The clear shift between the two histograms is a measurement
    of $dz/dt$ and therefore allows us to measure $H_0$.}
\label{agered}
\end{figure}

Figure~3 shows the derived single-burst equivalent ages of the
galaxies as a function of redshift.  The circles correspond to
galaxies in the SDSS LRG sample, triangles refer to the Treu et al.\
sample, diamonds are galaxies in MS1054$-$0321, and crosses are 53W091
and 53W069.  For clarity, the SDSS LRG points have not been plotted
for ages $<7$ Gyr.  Typical errors on the ages of LRG galaxies are
10\% and are not plotted.  An age--redshift relation (i.e., an
``edge'' or ``envelope'' of the galaxy distribution in the
age--redshift plane) is apparent from $z=0$ to $z=1.5$. \citet{JVTS03}
also show that small recent episodes of start formation do not
significantly bias the age determination.

What are the observational limits that can be imposed on $w(z)$? As a
consistency check we note that the age at $z=0$ obtained with this
method is in good agreement with GC ages and that the Treu et al.\
galaxy ages agree with those from of SDSS LRG sample where they
overlap.  The solid line in Figure~3 corresponds to the age--redshift
relation for a flat, reference $\Lambda$CDM model: $\Omega_m=0.27$,
$h=71$, and $w=-1$ (WMAP best fit model). This is consistent with the
observed age--redshift envelope, and it seems to indicate that we live
in a universe with a classical vacuum energy density and that, on
average, stars in the oldest galaxies formed about 0.7 Gyr after the
Big Bang (e.g., at redshifts $z \sim 7-10$). To make the observed
age--redshift relation consistent with a model with widely different
$w(z)$ behavior, one would have to infer that early-type galaxies at
different redshifts have very different formation epochs for their
stellar populations.  For example, the dotted line in Figure~3
indicates the age-redshift relation for a passively evolving
population in a universe where $w(z)=-2$ for $z > 1$ and grows
linearly from $w=-2$ at $z=1$ to $w=0$ at $z=0$.  For this model to
work, we would have to conclude that the oldest galaxies at $z\sim
1.5$ formed about 0.7 Gyr after the Big Bang, but that the oldest
galaxies at $z\sim 0$ formed $\sim 3$ Gyr after the Big Bang.  Since
the oldest galaxies in the LRG sample are as old as the oldest GCs, we
consider this to be an unlikely explanation.  Furthermore, not only
would this scenario require that {\em all} local galaxies either
formed their stars more recently than galaxies at high redshift or had
recent episodes of star formation, but would also require a remarkable
fine-tuning for them all to end up with exactly the same age. We
therefore conclude that this extreme model for $w(z)$ is unlikely
given the data.

Unfortunately, the small number of galaxies in \citet{JVTS03} samples
at $z>0.2$ did not allow them to compute $dz/dt$ with enough accuracy
to constrain $w(z)$, and therefore this measurement will have to await
better data.  However, the $z < 0.2$ region is well populated by LRG
galaxies (Figure~3) and they determined $dt/dz$ at $z\sim 0$. This is
illustrated in the bottom panel of Figure~3 by the clear shift in the
upper envelope of the age histogram between the two redshift ranges $0
< z < 0.04$ and $0.08 < z < 0.12$. This can be used to determine
$H_0$, which relies upon determining the ``edge'' of the galaxy age
distribution in different redshift intervals.

If the ages of the galaxies were known with infinite accuracy, for
each galaxy$_i$ at redshift $z_i$, one could associate an age $a_i$.
The probability for the age of the oldest stellar population in
galaxy$_i$, $P(t|a_i)$, would be given by a step function which jumps
from 0 to 1 at $a_i$.  An age--redshift relation edge could then be
obtained by dividing the galaxy sample into suitably-large redshift bins
and multiplying the $P(t|a_i)$ for all the galaxies in each bin.

\begin{figure}
\begin{center}
\includegraphics*[width=10cm]{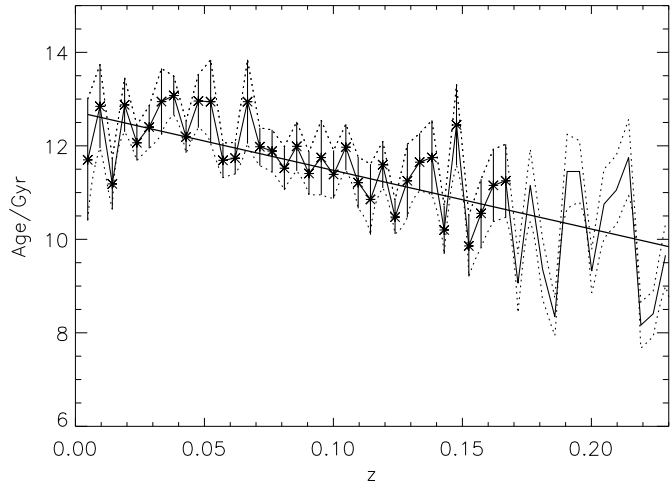}
\end{center} 
\caption{The binned, low-redshift age--redshift relation, derived
from the SDSS LRG sample.  The solid line is a best-fit to the edge.}
\label{fig3}
\end{figure}

In practice, the age of each galaxy is measured with some error
$\delta a_i$.  We thus assume $P(t|a_i)=1$ if $t>a_i+\delta_{a_i}$ and
$\ln P(a_i)=-x^2$ where
$x=(a_i+\delta_{a_i}-t)/(\sqrt{2}\delta_{a_i})$
otherwise. \citet{JVTS03} divided the $z < 0.2$ portion of the LRG
sample into 51 redshift bins.  For each bin, they obtained $P(z,t)$ by
multiplying the $P(t|a_i)$ of the galaxies in that bin.  They define the
``edge'' of the distribution $t(z)$ to be where $\ln P(z,t)$ drops by
$0.5$ from its maximum (i.e. $\Delta \ln P=0.5$ and associate an error
to this determination $\delta_{t(z)}$ given by $t_{2}(z)-t(z)$ where
$t_{2}(z)$ corresponds to where $\ln P(z,t)$ drops by $2$ from its
maximum (i.e. $\Delta \ln P=2$, this approximately corresponds to the
68\% confidence level).  This procedure makes the determination of the
``edge'' less sensitive to the outliers (see also
\citet{RBBG97}). Figure~4 presents the resulting $t(z)$ relation.

\begin{figure}
\begin{center}
\includegraphics*[width=10cm]{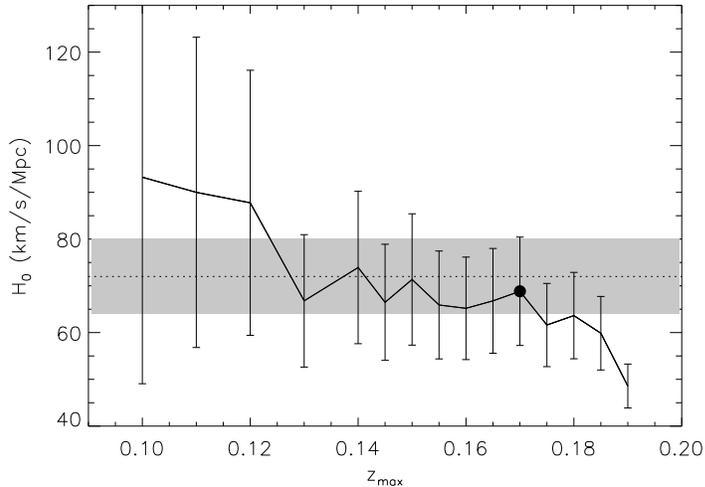}
\end{center} 
\caption{The value for $H_0$ derived from Figure~\ref{fig3} as a
    function of redshift cut-off for computing the slope. Note that it
    is very weakly dependent on the value of this cut-off. The solid
    circle is the adopted value. The shaded are corresponds to the
    1$\sigma$ confidence region from the Hubble Key Project. For
    $z>0.18$ the probability of the fit being a straight line is
    $P_{\ge \chi^2}<0.1$, while at $z=0.17$ is 0.32.}
\label{fig4}
\end{figure}

For $z<z_{\rm max}$, \citet{JVTS03} fit $t(z)$ with a straight line
whose slope $dt/dz$ is related to the Hubble constant at an effective
redshift by $H(z_{\rm eff})= - \frac{1}{1+z_{\rm eff}}
(\frac{dt}{dz})^{-1}$.  This fit is performed by standard $\chi^2$
minimization.  They also compute $P_{\ge \chi^2}$, the probability of
obtaining equal or greater value of the reduced $\chi^2$ if the $t(z)$
points were truly lying on a straight line.  Values of $P_{\ge
\chi^2}<0.1$ means that a straight line is not a good fit to the
points.

For $w = -1$, $H_0=H(z_{\rm eff})\left[\Omega_m(1+z_{\rm
eff})^3+\Omega_{\Lambda}\right]^{-\frac{1}{2}}$ where $\Omega_m=0.27$,
$\Omega_\Lambda=0.7$. Figure~5 shows how the $H_0$ measurement depends
on $z_{\rm max}$. For $z_{\rm max} > 0.17$ not only the value for
$H_0$ drifts but also a straight line is not a good fit to the points
($P_{\ge \chi^2}$ suddenly drops below $0.1$). For their $H_0$
determination, $z_{\rm max}=0.17$, $P_{\ge \chi^2}=0.32$, $z_{\rm
eff}=0.09$, and the correction from $H(z_{\rm eff})$ to $H_0$ is a
$4\%$ effect. They obtained $H_0=69\pm 12$ km s$^{-1}$
Mpc$^{-1}$. Which is in good agreement with the Hubble Key Project
measurement \citep[$h = 0.72 \pm 0.03 \pm 0.0 7$;][]{Freedman+01},
with the valued derived from the joint likelihood analysis of {\it
WMAP} + 2dFGRS + Lyman-$\alpha$ forest power spectrum \citep[$h =
0.71^{+0.04}_{-0.08}$;][]{SpergelWMAP03}, with gravitational lens time
delay determinations \citep[$h = 0.59^{+0.12}_{-0.07} \pm
0.03$;][]{TreuKoopmans02}, and Sunyaev-Zeldovich measurements
\citep[$h = 0.60^{+0.04+0.13}_{0.04-0.18}$;][]{Reese+02}.

This check at $z=0$ gives some confidence about the possibility of
measuring $w(z)$ when larger samples of elliptical galaxies at higher
redshifts become available in the near future. DEEP2 and the full SDSS
catalog will provide excellent datasets for doing this.

I warmly thank my collaborators Avi Loeb, Dan Stern, Tommaso Treu and
Licia Verde.


\end{document}